\documentclass{mem}
\usepackage{natbib}\usepackage{txfonts}\usepackage{balance}
\usepackage{graphicx}
\usepackage[a4paper]{hyperref}
\idline{00}{001}
\begin{document}
\def\kms{$\mathrm {km~s}^{-1}$}
\def\ms{$\mathrm {m~s}^{-1}$}
\def\etal{{\it et al.}}
\def\dmm{$\Delta\mu/\mu$}
\def\daa{$\Delta\alpha/\alpha$}
\def\zabs{$z_{\rm abs}$}

\title{
Spatial and temporal variations of fundamental constants
}

\subtitle{}

\author{
S. A. \,Levshakov\inst{1} 
\and I. I. \,Agafonova\inst{1}
\and P. \,Molaro\inst{2}
\and D. \,Reimers\inst{3}
          }

\offprints{S. A. Levshakov}

\institute{
Ioffe Physical-Technical Institute, 
St. Petersburg, Russia\\
\email{lev@astro.ioffe.rssi.ru}
\and
INAF~-- Osservatorio Astronomico di Trieste, 
Trieste, Italy
\and
Hamburger Sternwarte, Hamburg, Germany
}

\authorrunning{Levshakov \etal}

\titlerunning{Spatial and temporal variations of fundamental constants}

\abstract{
Spatial and temporal variations in the electron-to-proton mass ratio, $\mu$, and in the
fine-structure constant, $\alpha$, are predicted in non-Standard models aimed to explain
the nature of dark energy.
Among them the so-called chameleon-like scalar field models predict strong dependence of masses and
coupling constants on the local matter density. 
To explore such models
we estimated the parameters
$\Delta \mu/\mu \equiv (\mu_{\rm obs} - \mu_{\rm lab})/\mu_{\rm lab}$ 
and
$\Delta \alpha/\alpha \equiv (\alpha_{\rm obs} - \alpha_{\rm lab})/\alpha_{\rm lab}$ 
in two essentially different environments,~-- terrestrial (high density) and
interstellar (low density),~--
from radio astronomical observations of cold prestellar molecular cores 
in the disk of the Milky Way.  
We found that 
$\Delta\mu/\mu = (22\pm4_{\rm stat}\pm3_{\rm sys})\times10^{-9}$,
and
$|\Delta \alpha/\alpha| < 1.1\times10^{-7}$.
If only a conservative upper limit is considered, then 
$|\Delta\mu/\mu| \leq 3\times10^{-8}$.
We also reviewed and re-analyzed 
the available data on the cosmological variation of $\alpha$ obtained from Fe\,{\sc i} and Fe\,{\sc ii}
systems in optical spectra of quasars. 
We show that statistically significant
evidence for the changing $\alpha$ at the level of $10^{-6}$ has not been provided so far.
The most stringent constraint on
$|\Delta \alpha/\alpha| < 2\times10^{-6}$
was found from the Fe\,{\sc ii} system at $z = 1.15$ towards the bright quasar HE 0515--4414.
The limit of $2\times10^{-6}$ corresponds to the utmost accuracy which can be reached with
available to date optical facilities.
}
\maketitle{}

\section{Introduction}

Spatial and temporal variations in the electron-to-proton mass ratio, 
$\mu \equiv m_{\rm e}/m_{\rm p}$, 
and in the fine-structure constant, 
$\alpha \equiv e^2/(\hbar c)$, 
are not present in the Standard Model of particle physics but they arise
quite naturally in grant unification theories, multidimensional theories and in general
when a coupling of light scalar fields to baryonic matter is considered
\citep{uz03,mar08,ch09}.
The light scalar fields are usually attributed to a negative pressure substance permeating the
entire visible Universe and known as dark energy \citep{cds98}. 
This substance is thought to be responsible
for a cosmic acceleration at low redshifts, $z < 1$ \citep{pr03}.
However, scalar fields cause a problem since they could violate the equivalence principle
which has never been detected in local tests \citep{tu04}.
A plausible solution of this dissension
was suggested with a so-called `chameleon' model 
which assumes that a light scalar field acquires an effective potential and an effective mass
due to its coupling to matter that strongly depends on the ambient matter density 
\citep{kw04, ms07, op08}.  
In such a way the scalar field may evade local tests of the equivalence principle 
since the range of the scalar-mediated force is too short ($\lambda_{\rm eff} \la 1$ mm)
to be revealed at the terrestrial matter densities.
This is not the case, however, 
for the space based tests where the matter density is considerably lower,
an effective mass of the scalar field is negligible,
and an effective range for the scalar-mediated force is very large ($\lambda_{\rm eff} \ga 1$ pc).

Calculations of atomic and molecular spectra show that different
transitions have different sensitivities to changes in fundamental constants 
\citep{vl93,dz99,ko08}.
Thus, measuring 
relative radial velocities, $\Delta V$,  between such transitions one can probe the hypothetical
variation of physical constants. 
For instance, a spatial dependence of $\mu$ and $\alpha$ on the ambient matter density 
can be tested locally using terrestrial measurements of molecular transitions and 
comparing them with radio astronomical observations of molecular clouds in the disk of the Milky Way  
(here we have a typical difference between the environmental gas densities of 
$\sim$19 orders of magnitude).  
Complementary to these measurements, a temporal dependence of $\mu$ and $\alpha$ on cosmic time can be
tested through observations of high-redshift intervening absorbers seen in quasar spectra. 

In this presentation we summarize our recent results on spatial and temporal variations of 
$\mu$ and $\alpha$ obtained at high redshifts $z > 0$ (QSO absorption systems) and at
$z=0$ (the Milky Way disk).

\section{Cosmological tests ($z > 0$)}

The cosmological variability of $\alpha$ can be probed by 
different methods. One of them,~---  the many-multiplet
method,~--- was suggested by \cite{dz99} and extensively used in the analysis
of quasar metal absorbers \citep{W99,mwf03,C04}. 
This method
is based on the comparison of wavelengths
of different transitions in different ions having different sensitivity coefficients to
the variation of $\alpha$ and rising from the same QSO absorption-line system.
However, the approach to estimate \daa\ on base of different ions
seems not to be very favorable: it requires many absorption systems in order to
suppress kinematic effects caused by irregular velocity shifts between
different ions, and
its averaging procedure over many sight lines and over large range of redshifts
may smear out a putative weak signal in \daa.
It is clear that in order to obtain \daa\ estimates
at every particular redshift the kinematic effects must be suppressed.
Such a procedure can be realized if lines of only one ion and
arising from the same atomic levels are utilized \citep{lev04}.
The corresponding method,~--- the single ion differential $\alpha$ measurement (SIDAM),~---
and its applications to Fe\,{\sc i} and Fe\,{\sc ii} systems are
described below.

Fe\,{\sc i} and Fe\,{\sc ii} are represented in optical spectra of quasars
by several resonance transitions having different sensitivity coefficients.
Under these conditions the accuracy of the \daa\ estimates is limited
by only two factors: (1) uncertainties in the calibration of the wavelength scale
(see, e.g., Centuri\'on \etal, these proceedings),
and (2) unknown abundances of iron isotopes in a particular absorption system.
For quasar spectra taken with the VLT, it is possible to reach the
accuracy in the wavelength calibration of 30-50 \ms\ \citep{MLM08}
which translates into the error in \daa\ of $\sim$2$\times10^{-6}$.
The boundaries on isotopic composition of iron do not
exceed $10^{-6}$ \citep{kk04,pkr09}.
Thus, a single absorption system with iron ions can provide
the accuracy of the \daa\ estimate at the level of $10^{-6}$.

\subsection{Fe\,{\sc i} at \zabs\ = 0.45}

Many resonance transitions 
of neutral iron Fe\,{\sc i} associated with the absorption-line
system at \zabs\ = 0.45 were identified in the spectrum of HE~0001--2340
by \cite{VD07}.
Among them there are Fe\,{\sc i} lines with quite
different sensitivity coefficients which makes
such systems suitable for individual \daa\ estimates.
The sensitivity coefficients for Fe\,{\sc i} transitions were
calculated by \cite{DzF08} on our request and revealed to be highly sensitive 
to $\alpha$ change.

\begin{figure}[t!]
\resizebox{\hsize}{!}{\includegraphics[clip=true]{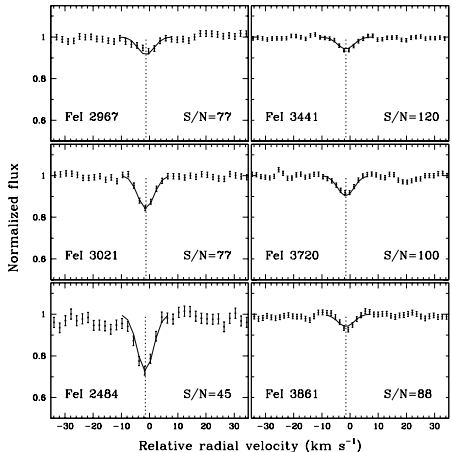}}
\vspace{-1.0cm}
\caption{\footnotesize
Combined absorption-line spectra of
Fe\,{\sc i} associated with the
\zabs = 0.45 absorption-line system towards HE~0001--2340
(normalized intensities are shown by dots with $1\sigma$ error bars).
The smooth curves show the model profiles.
The normalized $\chi^2_\nu = 0.80$ $(\nu = 59)$.
The zero radial velocity is fixed at $z = 0.45207$.
}
\label{fig1}
\end{figure}

\begin{figure}[t!]
\includegraphics[viewport=20 60 300 560,width=30mm,height=40mm]{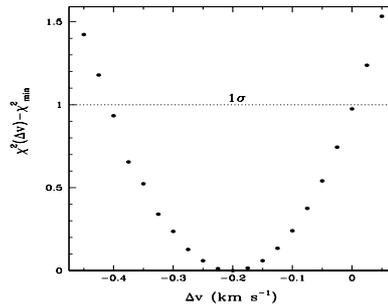}
\vspace{0.cm}
\caption{\footnotesize
$\chi^2$ as a function of the velocity difference $\Delta V$
between the Fe\,{\sc i} $\lambda\lambda$2967, 3021 \AA\ lines
and $\lambda\lambda$2484, 3441, 3720, 3861 \AA\
lines for the one-component model shown in Fig.~\ref{fig1}.
The $1\sigma$ confidence level is determined by $\Delta \chi^2 = 1$
which gives $\sigma_{\Delta v} = 200$ \ms.
}
\label{fig2}
\end{figure}

The profiles of the six selected Fe\,{\sc i} lines are shown in 
Fig.~\ref{fig1}.
In this case a simple one-component model can
describe adequately the observed intensities.
The most probable value of \daa\ = $(7\pm7)\times10^{-6}$ was calculated from
the $\Delta \chi^2$-curve presented in Fig.~\ref{fig2}.
Here we combined the Fe\,{\sc i} $\lambda\lambda$2967, 3021 \AA\ lines
having ${\cal Q} \simeq 0.08$ and the
Fe\,{\sc i} $\lambda\lambda$2484, 3441, 3720, 3861 \AA\ lines for which
${\cal Q} \simeq 0.03$. 
In Fig.~\ref{fig2}, the parameter $\Delta V$
is the velocity offset between these two groups of Fe\,{\sc i} lines.
In this case the estimate of \daa\ is given by
\begin{equation}
\Delta\alpha/\alpha =
\Delta V / (2c\Delta{\cal Q})\, ,
\label{eq5}
\end{equation}
where 
$\Delta V = V_1 - V_2$ is the difference between the radial
velocities of these two groups of lines, and
$\Delta {\cal Q} = {\cal Q}_2 - {\cal Q}_1$.

We expect that with higher spectral resolution and higher S/N 
the accuracy of the \daa\ estimate at \zabs\ = 0.45
can be increased by several times since the Fe\,{\sc i}
lines in this system are extremely narrow 
(the Doppler width $\la 1$ \kms)  
and have very simple profiles \citep{VD07}.

\begin{figure}[t!]
\includegraphics[viewport=138 200 418 715,width=60mm,height=80mm]{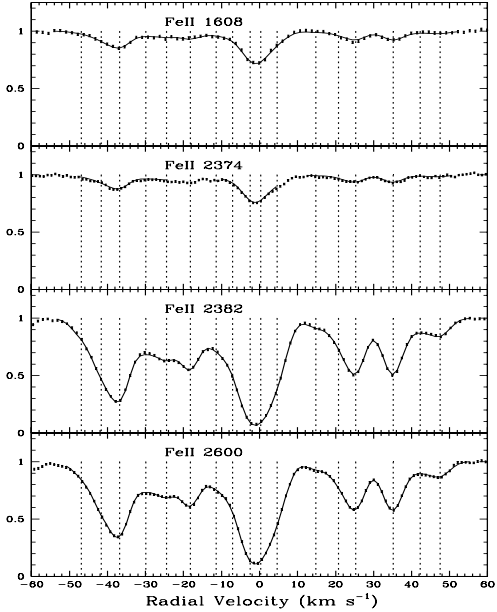}
\vspace{0.1cm}
\caption{\footnotesize
Profiles of Fe\,{\sc ii} lines associated with the
\zabs = 1.84 damped Ly$\alpha$ system towards Q 1101--264
(normalized intensities are shown by dots with $1\sigma$ error bars).
The $S/N$ ratios per pixel (from the top panel to the bottom) are 100,
138, 120, 120.
The smooth curves show the model profiles.
Parts of the Fe\,{\sc ii} $\lambda$2374 \AA\ line
are contaminated in the ranges $-25 < V -10$ \kms\ and $4 < V < 13$ \kms\
with telluric absorptions and not included in the analysis.
The positions of the
subcomponents of the equidispersion 17-component model
are marked by the dotted vertical lines.
The normalized $\chi^2_\nu = 0.87$ $(\nu = 300)$.
The zero radial velocity is fixed at $z = 1.838911$.
}
\label{fig3}
\end{figure}

\subsection{Fe\,{\sc ii} at \zabs\ = 1.84}

The only QSO spectrum which was obtained especially with the objective
to measure \daa\ is the VLT/UVES spectrum of a bright quasar
Q~1101--264 \citep{L07}.
This spectrum was observed with a high resolution 
(FWHM = 3.8 \kms) and S/N~$\ga 100$.
Analyzing the Fe\,{\sc ii} $\lambda\lambda$2600, 2382, and 1608 \AA\ lines
from the absorption system with \zabs\ = 1.84, we found
$\Delta V = -180\pm85$ \ms\ between
the Fe\,{\sc ii} $\lambda$1608 \AA\
line which has negative sensitivity coefficient
and the Fe\,{\sc ii} $\lambda\lambda$2600, 2382 \AA\ transitions having almost
identical positive sensitivity coefficients.
In terms of \daa\ this shift corresponds to \daa\ = $(5.4\pm2.5)\times10^{-6}$.

When new sensitivity coefficients for the Fe\,{\sc ii} lines
became available \citep{por07},
we repeated the analysis of this system including
the unblended parts of the Fe\,{\sc ii} $\lambda$2374 \AA\ line.
The profile of this line is blended
in the velocity range $-30 < V < -10$ \kms\ with telluric
absorption and because of that it was not considered in \cite{L07}.
This line has positive sensitivity coefficient
like the Fe\,{\sc ii} $\lambda\lambda$2600, 2382 \AA\ lines,
but its oscillator strength is close to that
of the weak line Fe\,{\sc ii} $\lambda$1608 \AA.
This allows us to control possible saturation effects in the strong lines
Fe\,{\sc ii} $\lambda\lambda$2600, 2382 \AA\
caused by unresolved components which may lead to some radial velocity
shifts between the apparent positions of the strong and weak Fe\,{\sc ii} lines.
The profiles of all four Fe\,{\sc ii} lines (Fig.~\ref{fig3}) were fitted
by a series of mathematical models differing both in the number of
components (from 8 to 17) and in the types of components:
either all components have individual $b$-parameters or
all components have identical $b$-parameters~--- the so-called equidispersion
deconvolution \citep{LTA99}.

\begin{figure}[t!]
\hspace{-1.0cm}
\includegraphics[viewport=0 170 260 670,width=30mm,height=40mm]{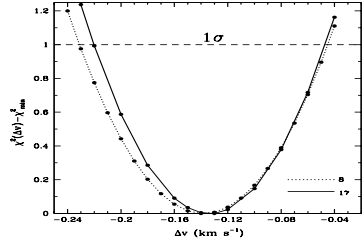}
\vspace{-0.7cm}
\caption{\footnotesize
$\chi^2$ as a function of the velocity difference $\Delta V$
between the Fe\,{\sc ii} $\lambda1608$ \AA\
and $\lambda\lambda2374, 2382, 2600$ \AA\
lines (shown in Fig.~\ref{fig3})
for the 8- and 17-component equidispersion models.
The $1\sigma$ confidence level is determined by $\Delta \chi^2 = 1$
which gives $\sigma_{\Delta v} = 90$ \ms\ for both represented models.
}
\label{fig4}
\end{figure}

\begin{figure}[t!]
\hspace{-1.0cm}
\resizebox{\hsize}{!}{\includegraphics[clip=true]{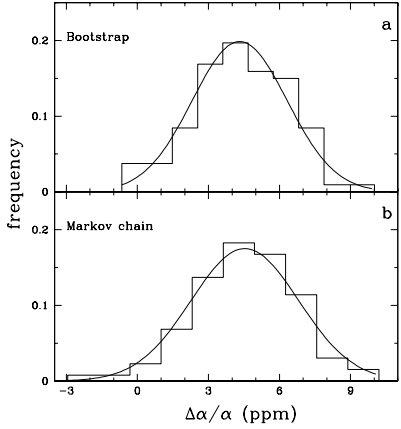}}
\vspace{-4.1cm}
\caption{\footnotesize
Histograms of the simulated samples generated by the
bootstrapping residuals (panel {\bf a}) and by the Markov chain
approximation (panel {\bf b}) for the 17-component
equidispersion model. The sample size $n = 100$ in both cases.
The smooth curves are Gaussians defined by the first two moments of
the corresponding samples: $(4.3\pm2.0)\times10^{-6}$ (panel {\bf a}), and
$(4.5\pm2.3)\times10^{-6}$ (panel {\bf b}).
}
\label{fig5}
\end{figure}

The 1$\sigma$ confidence limit for the velocity
shift between the Fe\,{\sc ii} $\lambda$1608 \AA\
and Fe\,{\sc ii} $\lambda\lambda$2600, 2383, 2374 \AA\
lines was calculated via the $\Delta \chi^2$-curve (Fig.~\ref{fig4})
and by means of Monte-Carlo simulations (Fig.~\ref{fig5}).
In turn, modeling of the data points in the Monte-Carlo simulations
occurred also in two ways: (1) the method of bootstrapping residuals, and
(2) the residuals were modeled by a Markov chain.
The bootstrap procedure breaks correlations between the
data points presented in the original QSO spectra
(see, e.g., Appendix A in Levshakov \etal\ 2002),
whereas the Markov chain approximation to the residuals allows us to save the
original correlation in the simulated data.
In all approaches the negative velocity shift between the Fe\,{\sc ii}
$\lambda$1608 \AA\ line
and the Fe\,{\sc ii} $\lambda\lambda$2600, 2382, 2374 \AA\ lines
was stable reproduced with the most probable value of
$\Delta V = -130\pm90$ \ms\
(as a 1$\sigma$ limit the most conservative value from all trials is taken).
In terms of \daa\ this corresponds to 
\daa\ = $(4.0\pm2.8)\times10^{-6}$.
We consider the result robust in this respect in disagreement 
with what claimed by \cite{mwf08}
who have not provided a quantitative analysis of the $z=1.84$ system
to support their claim.

\subsection{Fe\,{\sc ii} at \zabs\ = 1.15}

The most accurate to date 
estimate of \daa\ was obtained for the Fe\,{\sc ii} absorption system at \zabs = 1.15
detected in the spectrum of one of the
brightest high-redshift quasar HE 0515--4414 \citep{RH98}.
\cite{Q04} used co-added exposures taken with the VLT/UVES
and reported \daa\ = $(-0.4\pm1.9)\times10^{-6}$.
\cite{C06} re-observed this system with the high resolution and
temperature stabilized spectrograph HARPS mounted on the
ESO 3.6 m telescope at the La Silla observatory and obtained
\daa\ = $(0.5\pm2.4)\times10^{-6}$.
We note that in both publications
a wavelength calibration error was not taken into account, 
and once considered this their errors $\sigma_{\Delta\alpha/\alpha}$
would be a factor two larger.

Due to the brightness of HE 0515--4414, the $S/N$ ratio in individual
exposures of the VLT spectra is sufficiently high which allows us
to process them separately, i.e. without co-adding.
\cite{L06} analyzed the best quality individual exposures
and then averaged \daa\ values produced from the pairs of Fe\,{\sc ii}
lines (combinations of the blue line Fe\,{\sc ii} $\lambda$1608 \AA\
and different red Fe\,{\sc ii} lines).
The resulting mean value was \daa\ = $(-0.07\pm0.84)\times10^{-6}$.
However, it was overlooked that in this procedure the 
individual \daa\ values from different pairs
become correlated. Accounting for these correlations, 
as described in \cite{mol08},
and using the updated
sensitivity coefficients from \cite{por07}
we correct the latter \daa\ value to \daa\ = $(-0.12\pm1.79)\times10^{-6}$.
Thus, all \daa\ estimates at \zabs\ = 1.15, which are
the most accurate so far, exclude any cosmological
variations of $\alpha$ at the level of $2\times10^{-6}$.

\section{Local tests ($z=0$)}

\subsection{Estimate of \dmm} 

Among numerous molecules observed in the interstellar medium,
ammonia NH$_3$ is of a particular interest for the \dmm\ tests 
due to high sensitivity of the inversion frequency $(J,K) = (1,1)$ 
at 23.7 GHz to a change in $\mu$.
Here \dmm\ $\equiv (\mu_{\rm obs} - \mu_{\rm lab})/\mu_{\rm lab}$.
The sensitivity coefficient
(${\cal Q} = 4.5$) 
of the inversion transition was calculated
by \citep{fk07}:
\begin{equation}
\left({\Delta \nu}/{\nu} \right)_{\rm inv} \equiv
(\tilde{\nu} - \nu)/\nu  
= 4.5\ ({\Delta \mu}/{\mu})\ ,
\label{am1}
\end{equation}
where $\nu$ and $\tilde{\nu}$ are the frequencies corresponding to the
laboratory value of $\mu$ and to an altered  $\mu$ in a low-density
environment, respectively.
For the rotational frequencies we have
${\cal Q} = 1$ and
\begin{equation}
\left( {\Delta \nu}/{\nu} \right)_{\rm rot}
\equiv (\tilde{\nu} - \nu)/{\nu}
= {\Delta \mu}/{\mu}\ .
\label{am2}
\end{equation}
Comparing the
apparent radial velocities for the NH$_3$ inversion transition, $V_{\rm inv}$,
with rotational transitions, $V_{\rm rot}$, of other molecules 
arising {\it co-spatially} with NH$_3$
one finds
\begin{equation}
\Delta \mu/\mu = 0.3(V_{\rm rot} - V_{\rm inv})/{c}
\equiv 0.3{\Delta V}/{c}\ ,
\label{am4}
\end{equation}
where $c$ is the speed of light.

The parameter $\Delta V$ in (\ref{am4}) could be considered as  
the sum of two components,
$\Delta V = \Delta V_\mu + \Delta V_n$,
with $\Delta V_\mu$ being the shift due to $\mu$-variation, and $\Delta V_n$
is the so-called Doppler noise.
The input of the Doppler noise to a putative \dmm\ signal can
be reduced to some extent if the velocity shifts caused by
inhomogeneous distribution of molecules and 
instrumental imperfections are minimized
(for details, see Levshakov \etal\ 2009a). 

To probe \dmm\ under different local environments, 
we analyzed at first high resolution
spectra (FWHM = 25 \ms) of the NH$_3$ $(J,K) = (1,1)$ and CCS $J_N = 2_1 - 1_0$
transitions 
obtained with the 100-m Green Bank Telescope (GBT) by \cite{ros08} 
and \cite{rat08}, and moderate resolution spectra (FWHM = $120-500$ \ms)
of the NH$_3$ $(J,K) = (1,1)$, HC$_3$N $J=5-4$, and N$_2$H$^+$ $J=1-0$
transitions
observed with the 45-m Nobeyama radio telescope by \cite{sak08}.
These spectra showed a systematic velocity shift between
rotational and inversion transitions with the most accurate value
of $\Delta V = 52\pm7_{\rm stat}\pm13_{\rm sys}$ \ms\
based on the NH$_3$ and CCS lines \citep{lmk08}. 

Then we carried out our own observations of ammonia and other molecules with 
the 32-m Medicina, 100-m Effelsberg, and 45-m Nobeyama radio telescopes
\citep{lev09}.

Observations at the Medicina telescope were performed with both
available digital spectrometers ARCOS (ARcetri COrrelation Spectrometer)
and MSpec0 (high resolution digital spectrometer) which
have channel separations of 4.88~kHz and 2~kHz, respectively.
For ARCOS, this resolution corresponds to 62 \ms\
at the position of the ammonia inversion transition (23 GHz) and 80 \ms\
for the rotational HC$_3$N (2--1) line (18 GHz). For MSpec0, it is
25 \ms\  and 32 \ms\ at the corresponding frequencies.

The Effelsberg data were obtained 
with the HEMT (High Electron Mobility Transistor) dual channel receiver.
The measurements were obtained in a frequency switching mode,
with a frequency throw of $\sim$5\,MHz. The backend was an FFTS
(Fast Fourier Transform Spectrometer), operated with its minimum
bandwidth of 20\,MHz and simultaneously providing 16384 channels
for each polarization. The resulting channel widths are 15.4 and
20.1 \ms\ for NH$_3$ and HC$_3$N, respectively.

At Nobeyama we used a low-noise HEMT receiver, H22,
for the NH$_3$ observations and the two sideband-separating SIS
(Superconductor-Insulator-Superconductor)
receiver, T100 \citep{nak08}, for the N$_2$H$^+$ observations.
Both of them are dual polarization receivers. 
We observed two polarization simultaneously.  
Autocorrelators were employed as a backend with
bandwidth and channel separation 
of 4 MHz and 4.375 kHz, respectively.
This corresponds to the channel widths of 57 \ms\ at 23 GHz, and
14 \ms\ at 93 GHz.

\begin{figure}[t!]
\includegraphics[viewport=120 60 430 580,width=60mm,height=80mm]{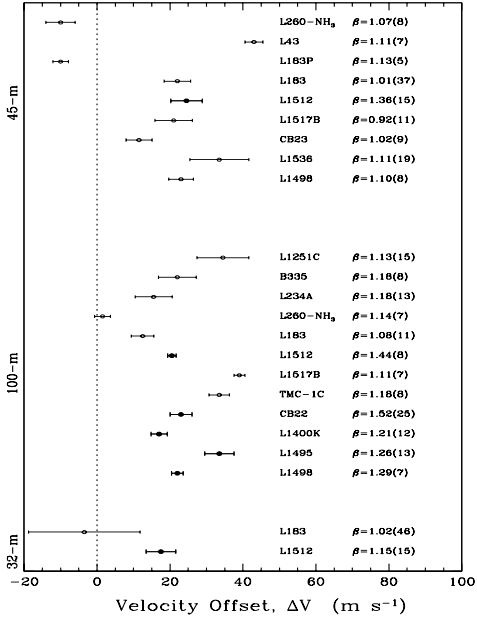}
\vspace{0.4cm}
\caption{\footnotesize
Radial velocity offsets, $\Delta V$, between the HC$_3$N $J=2-1$ and 
NH$_3$ $(J,K) = (1,1)$
transitions for the data obtained with the 32-m and 100-m telescopes, and between
the N$_2$H$^+$ $J=1-0$ and NH$_3$ $(J,K) = (1,1)$ transitions
for the 45-m telescope (1$\sigma$ statistical errors are indicated).
$\beta$ is the ratio of the Doppler $b$-parameters: $\beta = b$(NH$_3$)$/b$(HC$_3$N), or
$\beta = b$(NH$_3$)$/b$(N$_2$H$^+$). Given in parentheses are $1\sigma$ errors.
The filled circles mark sources with thermally dominated motions.
}
\label{fig0}
\end{figure}

The most accurate estimate in this case was based on the comparison
of the radial velocities of the NH$_3$ inversion line with
the rotational transition of HC$_3$N $J=2-1$ observed with
the 100-m Effelsberg telescope:
$\Delta V = 23\pm4_{\rm stat}\pm3_{\rm sys}$ \ms.
The distribution of radial velocity offsets between the rotational and
inversion transitions is shown in Fig.~\ref{fig0}.
The points represent molecular clouds with
symmetric line profiles which can be described by a single-component Gaussian model.
This requirement provides a better
accuracy of the line center measurements.
In the selected clouds the widths of the emission lines
do not exceed greatly the Doppler width due to thermal motion of material
(typical kinetic temperatures $T_{\rm kin} \sim 10$ K).
This ensures that turbulent motion does not
dominate in the line broadening and thus the selected molecular lines sample the same
kinetic temperature and arise most likely {\it co-spatially}.

The difference between the GBT and Effelsberg $\Delta V$ values
is probably connected to uncertainties in the laboratory frequencies.
The most accurate frequencies for the ammonia inversion transitions are
measured with the uncertainty $\varepsilon_v \la 0.6$ \ms\ \citep{ku67,ho72}. 
The HC$_3$N $J=2-1$ laboratory frequencies are known with the error 
$\varepsilon_v \simeq 3$ \ms\ \citep{mul05}.
The worst known is the frequency of the CCS $J_N = 2_1 - 1_0$
transition. The GBT radial velocities of CCS were measured with the
frequency $\nu = 22344.033(1)$ MHz estimated by \cite{yam90} from
radio astronomical observations (1 kHz at 22.3 GHz corresponds to 
$\varepsilon_v = 13.4$ \ms). 
The only laboratory measurement available gives $\nu = 22344.029(4)$
MHz \citep{lov92}. With this frequency we have 
$\Delta V = 6\pm7_{\rm stat}\pm54_{\rm sys}$ \ms.
The CDMS catalogue value $\nu = 22344.0308(10)$ MHz \citep{mul05}
yields $\Delta V = 22\pm7_{\rm stat}\pm13_{\rm sys}$ \ms\ which is in line with
the Effelsberg measurements based on the NH$_3$ and HC$_3$N lines.

These results show the importance of high precision laboratory
measurements which are expected to provide 
radio frequencies with uncertainties $\sim$1 \ms\ to be consistent with contemporary
radio astronomical observations. 

If we interpret the most accurate to date Effelsberg measurements of $\Delta V$
in terms of the electron-to-proton mass ratio variation, then 
\dmm\ $= (22\pm4_{\rm stat}\pm3_{\rm sys})\times10^{-9}$.

\subsection{Constraint on \daa} 

Complementary to the \dmm\ estimate, constraints on $\alpha$-variation can be
obtained from the comparison of the atomic fine-structure (FS) and molecular
rotational transitions \citep{ko08,lr08}.
In this case the velocity difference
between the rotational and FS lines is sensitive to the combination of
$\alpha$ and $\mu$, $F = \alpha^2/\mu$:
\begin{equation}
\Delta F/F \equiv 2\Delta \alpha/\alpha - \Delta \mu/\mu = \Delta V/c\, ,
\label{am5}
\end{equation}
where $\Delta V = V_{\rm rot} - V_{\rm fs}$.

For practical applications we may compare the FS transitions of carbon [C\,{\sc i}]
with rotational transitions of $^{13}$CO.
The spatial distribution of $^{13}$CO is closely
traced by the [{C}\,{\sc i}] FS lines \citep{sd97,io02,pa04}.
The rest hyper-fine frequencies of low-$J$ rotational
transitions of $^{13}$CO are known with very high accuracy,
$|\varepsilon_v| \la 0.1$ \ms\ \citep{cpl04}.
The most precise rest frequencies to date
for the [{C}\,{\sc i}] $J=1-0$ transition 492160.651(55) MHz \citep{YS91}
and $J=2-1$ transition 809341.97(5) MHz \citep{KLS98}
restrict the uncertainties of the line centers
to $\varepsilon_v = 33.5$ \ms\ and 18.5 \ms, respectively.
Thus, if we take $\varepsilon_v \simeq 34$ \ms\ as the most conservative
error of the line centering in a pair of $^{13}$CO--[\ion{C}{i}] lines,
then the limiting accuracy for $\Delta F/F$ is about $0.1$ ppm.

The analysis of the published results on [{C}\,{\sc i}] and
$^{13}$CO low resolution observations (FWHM = $0.2-1.0$ \kms)
of cold molecular cores gives
$\Delta V = 0\pm60_{\rm stat}\pm34_{\rm sys}$ \ms\ \citep{lmr09}, 
which leads to a limit
$|\Delta F/F| < 2.3\times10^{-7}$.
With the obtained estimate of \dmm, one finds from (\ref{am5}) that 
$|\Delta \alpha/\alpha| < 1.1\times10^{-7}$.

This result is already an order of magnitude more sensitive than
the bound on the cosmological $\alpha$-variation found from absorption
systems of quasars.
The present estimate of \daa\ at $z = 0$ can be considerably
improved if higher resolution spectra of molecular cores will be obtained and 
the rest frequencies of the [C\,{\sc i}] FS transitions will be measured with
better accuracy.

\section{Conclusions}

Our current results of astrophysical
tests on spatial and temporal variations
of fundamental constants can be summarized as follows.
\begin{enumerate}
\item[1.]
The relative radial velocities of 
the rotational transitions in HC$_3$N $(J=2-1)$ and the inversion
transition in NH$_3$ $(J,K)=(1,1)$ measured with the 100-m Effelsberg
radio telescope reveal a velocity offset of 
$\Delta V = 23\pm4_{\rm stat}\pm3_{\rm sys}$ \ms.
This offset is regularly reproduced in observations of different cold
molecular cores with different facilities at the 32-m Medicina and
45-m Nobeyama radio telescopes.
\item[2.]
Being interpreted in terms of the electron-to-proton mass ratio variation,
the found velocity offset corresponds to
\dmm\ = $(22\pm4_{\rm stat}\pm3_{\rm sys})\times10^{-9}$.
To cope with negative results obtained in laboratory experiments,
a chameleon-like scalar field is required to explain 
the non-zero $\Delta\mu$ value.
\item[3.]
A new approach to probe $\alpha$-variations using
the fine-structure transitions in atomic carbon [C\,{\sc i}]
and low-$J$ rotational transitions in $^{13}$CO is suggested,
and a strong constraint on the dependence of $\alpha$
on the ambient matter density is deduced at $z=0$:
$|\Delta\alpha/\alpha| < 1.1\times10^{-7}$.
\item[4.]
Another method to measure the cosmological variability of $\alpha$ from
resonance transitions of neutral iron Fe\,{\sc i} is discussed
and applied to the Fe\,{\sc i} system at \zabs\ = 0.45. The
resulting constraint is  
\daa\ = $(7\pm7)\times10^{-6}$. 
\item[5.]
The Fe\,{\sc ii} system at \zabs\ = 1.84 is re-analyzed and a
corrected value of \daa\ is obtained:
\daa\ = $(4.0\pm2.8)\times10^{-6}$.
\item[6.]
For the Fe\,{\sc ii} system at \zabs\ = 1.15 
the overlooked correlations between individual \daa\ values
are now taken into account that gives  
\daa\ = $(-0.12\pm1.79)\times10^{-6}$.
\item[7.]
It is shown that spectral radio observations
in the Milky Way are at present the most sensitive
tool to probe the values of fundamental constants at
different ambient physical conditions.
The internal (statistical) error of our Effelsberg data is
$\sigma_{\Delta\mu/\mu} = 4\times10^{-9}$ which is  
about 1000 times lower as compared with the error in
optical observations of quasars.
\item[8.]
Since extragalactic molecular clouds have
gas densities similar to those in the interstellar medium,
the value of
\dmm\ in high-$z$  molecular systems is expected to be at the same level as
in the Milky Way, i.e. \dmm$\sim$10$^{-8}$~--- providing
no temporal dependence of the electron-to-proton mass ratio is present.
\end{enumerate}

To conclude, we note that in order to be completely confident
that the revealed velocity shift between rotational and inversion lines
is not due to kinematic effects in the molecular
clouds but reflects a density-modulated variation of \dmm,
new high precision radio-astronomical
observations are needed for a wider range of objects. Besides,
such observations should include essentially different molecules 
with tunneling transitions sensitive to the changes in $\mu$ and
molecules with $\Lambda$-doublet lines which also exhibit enhanced
sensitivity to variations of $\mu$ and $\alpha$ (Kozlov 2009).
It is also very important to
measure the rest frequencies of molecular transitions
with an accuracy of about 1 \ms\ in laboratory experiments~---
in some cases the present uncertainties in the rest frequencies are larger
than the errors of radio-astronomical measurements. 
In addition, search for variations of the fine-structure
constant $\alpha$ in the Milky Way disk using mid- and
far-infrared fine-structure transitions in atoms and ions,
or search for variations of the combination of $\alpha^2/\mu$
using the [{C}\,{\sc ii}] and/or [{C}\,{\sc i}] fine-structure transitions and
low-lying rotational lines of CO 
would be of great importance for cross-checking the results.

\begin{acknowledgements}
The authors are grateful to the staffs of the Medicina,
Effelsberg, and Nobeyama radio observatories for 
technical assistance in observations.
The project has been supported by
DFG Sonderforschungsbereich SFB 676 Teilprojekt C4,
the RFBR grants 09-02-12223 and 09-02-00352,
and by the Federal Agency for Science and Innovations grant
NSh 2600.2008.2.
\end{acknowledgements}

\bibliographystyle{aa}

\end{document}